\documentclass[12pt]{iopart}

\usepackage{siunitx}

\begin{document}

\title[Comment]{Comment on "Comparison of six simulation codes for positive streamers in air"(Plasma Sources Sci. Technol. 27 (2018) 095002)}

\author{Xiang Li}

\address{School of Science, Beijing Information Science and Technology University}
\ead{lixbjemc@126.com}
\vspace{10pt}
\begin{indented}
\item[] \today
\end{indented}

\begin{abstract}
Recently, a comparison of six codes for streamer discharge simulations were performed in [1]. In this comment, we discuss about the big differences between the results obtained by the different codes using the same deterministic model, and raise questions on the convergence of the codes and the minimum spatial resolution that are required for a converged results. 
\end{abstract}

%
%
%
%
%

Numerical simulations play an important role in the study of streamer discharge physics, and there have been various numerical studies of streamer discharges during the last several decades [2-7]. Recently, a comparison of six codes for streamer discharge simulations were performed in [1] on both efficiency and accuracy, using "the same fluid model with the same transport coefficients". They concluded that "reasonably good agreement between models" had been reached "when sufficiently fine grids and corresponding small time steps" were used. Noting that two codes used in [1] were black-box COMSOL Multiphysics, our discussions only focus on the left four in-house codes.

As is a common sense, a reliable simulation should at least be mesh-independent and convergent, otherwise the results obtained from the simulations are meaningless or even cause misunderstanding to the physics. 

The first question is about the convergence of the results. Some results show in [1] can hardly be conclude to be "reasonably good agreement". For the first test case in [1], in Figs. 6 and 7, the maximal electric field by FR and CWI codes are roughly 130 and 160 kV/cm near $L(t)=0.9$ cm, respectively; for the second case, in Fig. 9, the electric field by ES and CWI codes are 200 and 40 kV/cm near $L(t)=0.9$ cm, and have an obviously different trend when $L(t)>0.7$ cm, which is too big a difference to be conclude to be "good agreement"; for the third case, the difference in the maximal electric field in Fig. 14 is roughly 10\% between CWI and ES codes. It is doubt whether these codes will really converge to the same results. It is possible that different models lead to a difference of, e.g., 10\%. However, we wish to emphasize that these results in [1] were numerically obtained using the same deterministic model and parameters; therefore, numerical convergence (between different meshes for the same code, and between different codes on a same fine enough mesh) is a minimum requirement. It is unreasonable to require all the codes to give exactly the same results, but a good example of "reasonable agreement" for streamer discharge simulation results by different codes that we can learn from, can be found in Figs. 11, 12 and 17 of [7]. A clear and accurate explanation would help to remove the confusion in [1] on the big difference among different codes for a same deterministic model, and convince the readers about the validity of these codes.

The second question is if the four codes converge, what is the minimum spatial and temporal resolution required by each code for a converged result. For the spatial discretization, the four codes use the same finite volume framework but with different slope limiters, which surely brings in different numerical viscosity. Therefore, the minimum spatial resolution requirement of each codes for a converged result could be different. Although it is always true that "sufficiently fine grids and corresponding small time steps" may lead to convergence, but it is still not clear what can be called "sufficiently fine". In [5], it was concluded that for cathode-direct and anode-direct streamers, 0.1-1 and 10 $\mu$m are the minimum spatial resolution requirements for convergence, respectively; while in [1], the minimum grid size used was typically larger than 1 $\mu$m. One would be curious to know which slope limiter allows the loosest spatial resolution requirement; and we believe such conclusion is beneficial to other scholars.

We hope these questions can be properly clarified to not only remove the confusions about the validity of these codes and the results generated by them, but also give other scholars more insights on the algorithm selection and simulation configuration.

\hspace{1cm}

[1] B Bagheri, J Teunissen, U Ebert, M M Becker, S. Chen, O Ducasse, O Eichwald, D Loffhagen,
A Luque, D Mihailova, J M Plewa, J van Dijk, and M Yousif. Comparison of six simulation
codes for positive streamers in air. Plasma Sources Sci. Technol, 27:095002, 2018.

[2] G E Georghiou, R Morrow, and A C Metaxas. A two dimensional, finite-element, flux-corrected
transport algorithm for the solution of gas discharge problems. Journal of Physics D:Applied
Physics, 33:2453–2466, 2000.

[3] S K Dhali and P F Williams. Two dimensional studies of streamers in gases. Journal of Applied
Physics, 62:4696–4707, 1987.

[4] A. A. Villa, L. Barbieri, M. Gondola, A.R. Leon-Garzon, and Malgesini R. An efficient algorithm
for corona simulation with complex chemical models. J. Comput. Phys., 337:233–251, 2017.

[5] Pancheshnyi S., P. Segur, P. Capeilere, and A. Bourdon. Numerical simulation of filamentary discharges with parallel adaptive mesh refinement. Journal of Computational Physics, 227:6574–6590, 2008.

[6] B. Lin, C. Zhuang, Z. Cai, R. Zeng, and W. Bao. An efficient and accurate mpi-based parallel
simulator for streamer discharges in three dimensions. Journal of Computational Physics,
401:109026, 2020.

[7] I. L. Semenov and K.-D. Weltmann. A spectral element method for modelling streamer discharges in low-temperature atmospheric-pressure plasmas. arXiv:2111.06210, 2022.
\end{document}